# Microscopic dynamics perspective on the relationship between Poisson's ratio and ductility of metallic glasses

K. L. Ngai[1,2*], Li-Min Wang[2], Riping Liu[2], W. H. Wang[3]

[1]*Dipartimento di Fisica, Università di Pisa, Largo B. Pontecorvo 3, I-56127, Pisa, Italy*

[2]*State Key Lab of Metastable Materials Science and Technology, Yanshan University, Qinhuangdao, Hebei, 066004 People's Republic of China*

[3] *Institute of Physics, Chinese Academy of Sciences, Beijing 100190, People's Republic of China*

**Abstract**

In metallic glasses a clear correlation had been established between plasticity or ductility with the Poisson's ratio $\nu_{Poisson}$ and alternatively the ratio of the elastic bulk modulus to the shear modulus, $K/G$. Such a correlation between these two macroscopic mechanical properties is intriguing and is challenging to explain from the dynamics on a microscopic level. A recent experimental study has found a connection of ductility to the secondary β-relaxation in metallic glasses. The strain rate and temperature dependencies of the ductile-brittle transition are similar to the reciprocal of the secondary $\beta$-relaxation time, $\tau_\beta$. Moreover, metallic glass is more ductile if the relaxation strength of the $\beta$-relaxation is larger and $\tau_\beta$ is shorter. The findings indicate the $\beta$-relaxation is related to and instrumental for ductility. On the other hand, $K/G$ or $\nu_{Poisson}$ is related to the effective Debye-Waller factor (*i.e.*, the non-ergodicity parameter), $f_0$, characterizing the dynamics of a structural unit inside a cage formed by other units, and manifested as the nearly constant loss shown in the frequency dependent susceptibility. We make the connection of $f_0$ to the non-exponentiality parameter $n$ in the Kohlrausch stretched exponential correlation function of the structural α-relaxation function, $\phi(t) = \exp\left[-\left(\frac{t}{\tau_\alpha}\right)^{1-n}\right]$. This connection follows from the fact that both $f_0$ and $n$ are determined by the inter-particle potential, and $1/f_0$ or $(1-f_0)$ and $n$ both increase with anharmonicity of the potential. A well tested result from the Coupling Model is used to show that $\tau_\beta$ is completely determined by $\tau_\alpha$ and $n$. From the string of relations, (*i*) $K/G$ or $\nu_{\text{Poisson}}$ with $1/f_0$ or $(1-f_0)$, (*ii*) $1/f_0$ or $(1-f_0)$ with $n$, and (*iii*) $\tau_\alpha$ and $n$ with $\tau_\beta$, we arrive at the desired relation between $K/G$ or $\nu_{\text{Poisson}}$ and $\tau_\beta$. On combining this relation with that between ductility and $\tau_\beta$, we have finally an explanation of the empirical correlation between ductility and the Poisson's ratio $\nu_{\text{Poisson}}$ or $K/G$ based on microscopic dynamical properties.

*Author to whom correspondence should be addressed: ngai@df.unipi.it

## I. Introduction

The establishment of correlations between elastic constants and plastic deformation or ductility in metallic glasses of diverse compositions has been a major endeavor of recent years. The potential benefits of the correlations include better understanding of the plastic deformation and fracture mechanisms, the embrittlement by annealing, and in providing a strategy in the design of materials with improved performance in applications. The fact that ductile metallic glasses tend to have a large Poisson's ratio $\nu_{\text{Poisson}}$ or larger ratio of the elastic bulk modulus $K$ to the shear modulus $G$ was first reported in 1975 by Chen et al. [1]. But it is not until some thirty years later that the correlation becomes well defined by the study of a variety of bulk metallic glasses with different compositions [2,3]. The measured energy of fracture $G_c$ (i.e., the energy required to create two new surfaces by fracture) of as-cast metallic glasses showed a clear correlation with the elastic modulus ratio $K/G$. Moreover, ductile and brittle bulk metallic glasses (BMGs) seem to be separated by values of $G/K$ smaller or larger than 0.41–0.43 respectively, or equivalently with $\nu_{Poisson}$ larger or smaller than 0.31–0.32 respectively. The platinum-based glasses with high $K/G$ have exceptionally high toughness and malleability [4] also conform well to the correlation [2].

Annealing of metallic glasses is known to induce embrittlement, as demonstrated in [2,5] by the correlation of fracture energy $G_c$ with the ratio $K/G$ as the BMG $Zr_{41}Ti_{14}Cu_{12.5}Ni_{10}Be_{22.5}$ (Vitreloy 1) is annealed up to 24 hours at 623 K. The correlation turns out to be the same as that found when comparing different BMGs. The transition from ductility/plasticity to brittleness on annealing occurs at the same critical value of $K/G$ as found in the as-cast glasses of different compositions. These properties were demonstrated by putting all the data of the annealed Vitreloy 1 together with that of as-cast BMG in the same plot of $G_c$ versus $G/K$ [2].

The correlation of ductility with $K/G$ or $\nu_{Poisson}$ is supported and strengthened by data in other BMGs [6-9], and the collection of results from more references than cited here can be found in the review published in 2012 [9], where more references on this subject are given. Thus the correlation seems to be universally observed in BMGs, and deserves explanation.

In parallel with the development of the correlation discussed above are the revelation from various experimental studies that the secondary or $\beta$-relaxation plays nontrivial roles in the observed dynamic properties of BMGs, and also its characteristic bear remarkable resemblance to that of the mechanically deformed BMGs. For dynamic properties, we can give the following experimental evidences. The $\beta$-relaxation is instrumental for the crystallization of BMG at temperatures below the glass transition temperature $T_g$ [10]. The properties of the $\beta$-relaxation are closely correlated with the fragility [11], and the degree of non-exponentiality [12] of the alloys in the supercooled liquid state. The diffusion motion of the smallest constituent atom in BMG has about the same activation energy as the $\beta$-relaxation [13]. From experiments on mechanical deformation of BMG we have the relation to the $\beta$-relaxation from the following observations.

(*A*) The activation of shear transformation zones (STZs) and $\beta$-relaxations in metallic glasses are directly related, with the activation energy of the $\beta$-relaxation nearly the same as the potential-energy barriers of STZs [14].

(*B*) The $\beta$-relaxation of $La_{68.5}Ni_{16}Al_{14}Co_{1.5}$ follows similar time-temperature dependence as the activation of the structural units of plastic deformations and global plasticity. The ductile–brittle transition strain rate is comparable to the frequency, $1/\tau_\beta$, of the $\beta$-relaxation, and has the same Arrhenius temperature dependence [15]. Plastic flow of BMG occurs through cooperative shearing of unstable STZs activated by shear stress [16]. Thus, on combining this property with (A), the connection of $\beta$-relaxation to plasticity seems clear.

(*C*) While $La_{68.5}Ni_{16}Al_{14}Co_{1.5}$ is ductile, other BMG such as Ce-based ones with similar $T_g$ are brittle even at the test temperature of $0.92T_g$ and the strain rate of $10^{-4}$ s$^{-1}$ [17]. It is noteworthy that the $\beta$-relaxation of $La_{68.5}Ni_{16}Al_{14}Co_{1.5}$ is fast and has large relaxation strength, by contrast the $\beta$-relaxation of Ce-based BMG is slow and with small relaxation strength. BMG having faster $\beta$-relaxation with larger amplitude can respond to deformation at higher rate, and naturally the BMG is more ductile. Such correlation between ductility and $\beta$-relaxation had been found in glassy polymers [18-20]. An example is the exceptionally ductile bisphenol-A polycarbonate (BPA-PC), which has a very fast and prominent $\beta$-relaxation. Tetramethyl BPA-PC (TMBPA-PC) having four additional methyl groups present on the two phenyl rings is very similar in structure to BPA-PC. The $\beta$-relaxation of TMBPA-PC is much slower and the polymer is less ductile than BPA-PC.

(*D*) It was found that the STZ volumes of the six BMG systems increase with Poisson's ratio and BMG ductility [21].

From the connection the $\beta$-relaxation has with ductile-brittle transition stated in (*B*) and ductility in (*C*), and with STZ stated in (*A*) and (*D*), the implication can be drawn that the $\beta$-relaxation has impact on ductility and hence possible relation to Poisson's ratio $\nu_{\text{Poisson}}$ via the empirical correlation found between ductility and $\nu_{\text{Poisson}}$.

In the sections to follow we shall show these two apparently universal phenomena, ductility with *K/G* (or $\nu_{\text{Poisson}}$) and ductility with the $\beta$-relaxation, are related. Both are stemming from the general physics of the dynamics of many-body interacting systems including BMG as a special case. Thus, the two phenomena are linked and can be explained simultaneously by means of the Coupling Model [22-25], which was constructed to deal with the dynamics of many-body interacting systems. First we shall follow previous work [26] to relate the *K/G* (or $\nu_{\text{Poisson}}$) to the effective Debye-Waller factor, $f_o$, of the glass. In the Mode Coupling Theory (MCT), $f_o$ is called the non-ergodicity parameter [27]. However, for $f_o$ we shall continue to use the term, effective Debye-Waller factor, since the present paper has no direct connection to MCT. Next we show anharmoncity of the inter-particle potential determines simultaneously $f_o$ and the degree of non-exponentiality of the structural α-relaxation represented by the coupling parameter, *n*, of the Coupling Model [22-25]. Combining the relation between *K/G* (or $\nu_{\text{Poisson}}$) and $f_o$ on the one hand and the relation between $f_o$ and *n* on the other hand, we can see that *K/G* (or $\nu_{\text{Poisson}}$) is related to or determined by *n*. Moreover, in general, glass-formers with larger *n* have shorter $\tau_\beta$ [25,28], shown recently also to be the case for BMG [29]. We have seen from property (*B*) that the *β*-relaxation is responsible for the ductile-brittle transition. Glasses are more ductile if the β-relaxation has its relaxation time $\tau_\beta$ shorter and its strength $\Delta_\beta$ stronger because this favorable condition renders the *β*-relaxation more facile to respond to deformation at a higher rate without brittle failure. From this direct and transparent relation between ductility and $\tau_\beta$, together with the dependence of $\tau_\beta$ on *n* and the relation established between *n* and *K/G* (or $\nu_{\text{Poisson}}$), we have finally arrived at a microscopic explanation of the empirical correlation between ductility and *K/G* (or $\nu_{\text{Poisson}}$).

## II. Relating *K/G* (or $\nu_{\text{Poisson}}$) to the effective Debye-Waller factor $f_o$

This task has essentially been done by Novikov and Sokolov [26]. Here we supply the steps in obtaining the relation between *K/G* (or $\nu_{\text{Poisson}}$) to the effective Debye-Waller factor $f_o$, not given in [26]. We start from the normalized density correlation function, $\phi_q(t) = <\rho_q(t)\rho_q(0)>/S_q$, where $\rho_q$ denotes density fluctuations at wave vector *q*, and $S_q = <(\rho_q)^2>$ is the static structure factor [27]. The decay of $\phi_q(t)$ occurs first by vibrations and then levels off to a plateau value, $f_q$, at longer times, but is terminated eventually by the onset of relaxation process, *β* or α [30]. Sound waves are determined by the density of the glass and the elastic modulus [31]. Longitudinal sound velocity v is determined from the longitudinal modulus $M_L(\omega)$ and density $\rho$ by $v = \sqrt{M_L/\rho}$. It is directly related to the long wavelength limit of the dynamical structure factor, $S(q,\omega)$ [26,27]. The last quantity is the Fourier transform of $<\rho_q(t)\rho_q(0)>$, and it can be expressed in terms of $M_L(\omega)$. Further development given in [26] leads to the following expression for the zero wave vector Debye-Waller factor,

$$f_0 = 1 - \frac{v_0^2}{v_\infty^2} = 1 - \frac{M_0}{M_\infty}, \tag{1}$$

where $v_0$ ($M_0$) and $v_\infty$ ($M_\infty$) are the zero and infinite frequency longitudinal sound velocity (modulus) respectively. Using the identity, $M = K + \left(\frac{4}{3}\right)G$, the expression for $f_0$ can be rewritten as

$$f_0 = 1 - [K_0 + (4/3)G_0]/[K_\infty + G_\infty] \tag{2}$$

From here we follow [25] in making approximations in the glassy state by putting the zero frequency shear modulus $G_0=0$, and neglecting the softening of the bulk modulus on decreasing frequency, i.e., putting $K_0 \cong K_\infty$. The final result on the relation between $f_0$ and $K_\infty/G_\infty$ is given by

$$f_0 = \frac{1}{\left(\frac{3}{4}\right)\left(\frac{K_\infty}{G_\infty}\right)+1} \tag{3}$$

Thus, BMG with larger $K/G$ has a smaller effective Debye-Waller factor $f_0$. In other words, $K/G$ correlates with $1/f_0$, and also $v_{\text{Poisson}}$ with $1/f_0$, because $v_{Poisson} = \frac{1}{2} - \frac{3}{(6K_\infty/G_\infty)+2}$.

### III. $1/f_0$ or $(1-f_0)$ correlates with $n$

We shall use molecular dynamics simulations as well as experimental data to support the correlation of $1/f_0$ or $(1-f_0)$ with non-exponentiality of the structural $\infty$-relaxation. The origin of this correlation can be traced back to both quantities are parallel consequences of the many-body dynamics governed by anharmonicity of the inter-particle potential. By the term non-exponentiality we mean the degree of departure of the correlation from linear exponential relaxation, and is measured by the parameter $n$ appearing in the Kohlrausch stretched exponential function,

$$\phi(t) = \exp\left[-(t/\tau_\alpha)^{1-n}\right]. \tag{4}$$

Eq.(4) is commonly used to fit the time dependence of the structural α-relaxation. Molecular dynamics simulations of binary Lennard-Jones (LJ) systems consisting of two kinds of particles A and B were performed [33,34] in which the potential, $V(r)$, was varied systematically to increase its anharmonicity and to observe the change in the dynamics. Three different LJ potentials

$$V(r) = \frac{E_0}{(q-p)}[p(r_0/r)^q - q(r_0/r)^p] \tag{5}$$

with ($q$=8, $p$=5), ($q$=12, $p$=6), and ($q$=12, $p$=11) were studied. The well-depth and the location of the minimum of $V(r)$ are the same for all three potentials. The ($q$=12, $p$=11) LJ potential is more harmonic than the classical ($q$=12, $p$=6) LJ potential, while the ($q$=8, $p$=5) LJ potential is a flat well and exceedingly anharmonic. The models using the ($q$=12, $p$=11), ($q$=12, $p$=6) and ($q$=8, $p$=5) potentials were referred to, in order of increasing anharmonicity, as Model I, II and III, respectively. The self-intermediate scattering function, $F_s(Q_0,t)$ for the A particles were calculated for $Q_0=2\pi/r_0$, where $r_0$ is the position of the maximum of the first peak of the static A-A pair correlation function. The $F_s(Q_0,t)$ exhibit a two-step decay separated by a plateau with height $f(Q_0,T)$, which is the effective Debye-Waller factor at wave vector $Q_0$. At the end of plateau, the decay of $F_s(Q_0,t)$ at temperature $T$ down to zero is approximately described by $F_s(Q,t)=f(Q_0,T)\exp[-(t/\tau_\alpha)^{1-n}]$. The β-relaxation was not resolved in these simulations possibly because it has weak strength and may have merged with the α-relaxation. To compare the dynamics of I, II, and III, temperature in each case is scaled by $T_{ref}$ at which $\tau_\alpha$ is equal to an arbitrarily chosen long time, The results of the simulation essential to the present paper can be summarized as follows. For all $T/T_{ref}$, increasing anharmonicity of the interparticle potential is accompanied by (*i*) increasing non-exponentiality of the α-relaxation or the coupling parameter $n(T/T_{ref})$ of the Coupling Model [22-25], (*ii*) decreasing effective Debye-Waller factor $f(Q_0, T/T_{ref})$ or equivalently increasing value of $1/f(Q_0,T/T_{ref})$ or $(1-f_0)$, (*iii*) increasing steepness index or fragility parameter, *m* [34]. Thus, correlations were established between four quantities, anharmonicity, *n*, $1/f(Q_0,T)$, and *m*, in the supercooled liquid state. In the glassy state where $f(Q_0,T)$ is contributed entirely from vibrations, the procedure of Scopigno et al. [35] had been used to obtain the zero wave vector limit of $f(Q\to 0, T/T_{ref})=f_0(T/T_{ref})$. The results of the simulations again show $1/f_0(T/T_{ref})$ is largest for case III, intermediate for case II, and smallest for case I. Hence, we have in the glassy state as well the correlations between anharmonicity, *n*, $1/f_0(T/T_{ref})$ or $(1-f_0)$, and *m* [33,34].

The increase of *n* with anharmonicity of the interparticle potential found by simulations is consistent with the expectation from the Coupling Model (CM) based on nonlinear Hamiltonian dynamics (i.e., classical chaos) of systems governed by anharmonic interaction potential [22-25]. Exact solutions of simplified model systems have demonstrated the increase of *n* with anharmonicty or nonlinearlity of the potential [23,24].

At the plateau level $f(Q\to 0,T)=f_0(T)$, where $F_s(Q\to 0,t)$ decreases very slowly with time and the corresponding mean square displacement $<r^2(t)>$ increases in proportional to a power law, $t^c$, with $c\ll 1$ [25,30]. In this regime, all particles are mutually caged via the anharmonic potential. By Fourier transforming $F_s(Q\to 0,t)$ to frequency space, the plateau becomes the imaginary part of the complex susceptibility $\chi''(\omega,T)=A(T)\omega^{-c}$ with $c\ll 1$ and the strength $A(T)$ weakly increases with temperature. This $\chi''(\omega,T)$ is often called the nearly constant loss (NCL) because $c\ll 1$. Larger value of $1/f_0(T)$ or $(1-f_0)$ corresponds to higher NCL or larger $A(T)$. Since the nearly constant loss (NCL) originates from particles mutually caged via the anharmonic inter-particle potential, we are led to expect that more anharmonic potential will give rise to larger NCL. Together with the proportionality relation, NCL $\propto 1/f_0(T)$ or $(1-f_0)$, this expectation leads to the result that larger value of $1/f_0(T)$ is associated with the more anharmonic potential. Combining this result with the already established correlation of anharmonicity with $n$ from the simulations and the CM, we arrive at the correlation between $n$ and $1/f_0(T)$ via the NCL by a different route.

Experimentally, NCL is found in all glass-formers at low enough temperature and/or higher frequencies by the dielectric loss spectroscopy [25,36,37], and also by the mean square displacement, $<r^2>$ observed at temperatures below $T_g$ in quasielastic neutron scattering experiments [30,38]. For metallic glasses, the presence of NCL was demonstrated by the weak temperature dependence of the isochronal shear mechanical loss, $G''(T)$, at low temperatures [39-42]. Larger NCL or $A(T)$ is found in glass-formers with larger $n$ from dielectric loss data [36], and from neutron scattering spectroscopy [42]. Again together with NCL $\propto 1/f_0(T)$ or $(1-f_0)$, there are plenty of support from experiments of the correlation between $1/f_0(T)$ or $(1-f_0)$ and $n$.

## IV. Correlation of $\tau_\alpha/\tau_\beta$ with $n$ for fixed $\tau_\alpha$

At any fixed value of $\tau_\alpha$, it has been verified repeatedly by experiments in different classes of glass-formers and simulations that $\tau_\beta$ of a special kind of secondary relaxation [25,28,43] is shorter for glass-former with larger $n$. Examples showing this property, plus more remarkable relations between $\tau_\beta$ and $\tau_\alpha$ no need to mention in this paper, can be found in a perspective [44] and in the book [25]. This special kind of secondary relaxation is often referred to as the Johari-Goldstein β-relaxation with the purpose of distinguishing it from other secondary relaxations which bear no relation to the α-relaxation. Metallic glasses has only one secondary relaxation either resolved or hidden under the high frequency flank of the α-loss peak. Hence it is the special kind of secondary relaxation, and the correlation of $\tau_\alpha/\tau_\beta$ with $n$ for fixed $\tau_\alpha$ was indeed found in metallic glasses [29].

## V. Correlation of ductility or *K/G* (or $\nu_{Poisson}$) with *n*, and with $\tau_\alpha/\tau_\beta$ for a fixed value of $\tau_\alpha$

In the Introduction we have mentioned an experiment demonstrating the connection between the *β*-relaxation and the activation of the structural units of plastic deformations and global plasticity [15]. The experiment was performed on $La_{68.5}Ni_{16}Al_{14}Co_{1.5}$ which shows unusually short $\tau_\beta$ or large $\tau_\alpha/\tau_\beta$, and also high macroscopic tensile plasticity. The ductile-brittle transition in tension follows similar time-temperature scaling relationship of the activation of the *β*-relaxation with energy $E_\beta$. The results obviously have the strong implications that the activation of the *β*-relaxation is responsible for the remarkable deformability observed in this BMG. That is, the more facile the *β*-relaxation, or shorter $\tau_\beta$ for a fixed value of $\tau_\alpha$, more ductile is the BMG at the same strain rate of deformation. This finding by itself is sufficient to support the correlation of $\tau_\beta$ with ductility.

If dynamic mechanical experimental data of *n* and/or $\tau_\beta$ were available together with *K/G* (or ν) or measure of ductility for the same BMG, the correlation between these two quantities or lack of it can be checked. Despite *K/G* (or $\nu_{Poisson}$) have been determined for a large number of BMG, dynamic mechanical data of *n*, $\tau_\beta$ and the ratio $\tau_\alpha/\tau_\beta$ for a fixed value of $\tau_\alpha$ for the same BMG are few. In some cases, the best one can find are data of *K/G* (or $\nu_{Poisson}$) and *n* or $\tau_\alpha/\tau_\beta$ in two BMG with nearly the same chemical composition. An example is $La_{70}Ni_{15}Al_{15}$ which is slightly different from $La_{68.5}Ni_{16}Al_{14}Co_{1.5}$ discussed in the above. If we can treat the two BMG as equivalent in properties, then the correlation of high tensile plasticity from $La_{68.5}Ni_{16}Al_{14}Co_{1.5}$ with very large $n=0.58\pm0.04$, and very short $\tau_\beta(T_g)=2\times10^{-5}$s for a fixed value of $\tau_\alpha(T_g)=10^3$ s from $La_{70}Ni_{15}Al_{15}$ [29] is obtained. It is fortunate that for $Pd_{40}Ni_{40}P_{20}$, and $Zr_{46.75}Ti_{8.25}Cu_{7.5}Ni_{10}Be_{27.5}$ (VIT4), all the parameters, $K/G$, $\nu_{Poisson}$, $n$, and $\tau_\beta(T_g)$ all are known. For $Pd_{40}Ni_{40}P_{20}$, we have $K/G=4.79$, $\nu=0.403$ [9], $n=0.50$, and $\tau_\beta(T_g)=2.3\times10^{-4}$ s [29], while for VIT4, $K/G=3.1$, $\nu_{Poisson}=0.356$ [9], $n=0.43$, and $\tau_\beta(T_g)=5.0\times10^{-4}$ s [29]. Both are ductile BMG like $La_{70}Ni_{15}Al_{15}$, and the parameters of the three BMG support the correlation of $K/G$ and $\nu_{Poisson}$ with $n$ and $1/\tau_\beta(T_g)$.

The pair, $Ce_{68}Al_{10}Cu_{20}Nb_2$ and $Ce_{70}Al_{10}Cu_{20}$, having slight difference in composition by being with and without the small amount of Nb. They have mechanical properties in stark contrast to the ductile BMG discussed above. The Ce-based BMG with comparable $T_g$ as the La-based BMG are brittle even under the test temperature of $0.92T_g$ and the strain rate of $10^{-4}$ s$^{-1}$ [45]. Obeying the correlation between ductility and elastic constants, brittle $Ce_{68}Al_{10}Cu_{20}Nb_2$ has the small values of $K/G=2.5$ and $\nu_{Poisson}=0.32$ [9]. The dynamic mechanical relaxation study of $Ce_{70}Al_{10}Cu_{20}$ yields a small value of $n=0.20$, and long $\tau_\beta(T_g)$ estimated to be about 1 s [46]. If we can treat the pair of Ce-based BMG as one and the same material, the smaller values of $K/G=2.5$ and $\nu_{Poisson}=0.32$ correlates with small values of $n=0.20$, and long $\tau_\beta(T_g)\approx1$ s.

X-ray photon correlation spectroscopy was employed to investigate the structural relaxation process in the metallic glass, $Mg_{65}Cu_{25}Y_{10}$, [47] on the atomic length scale. In the supercooled liquid state, the Kohlrausch stretched exponential fits to the correlation function yield $(1-n)=0.88$, or $n=0.12$. The Poisson's ratio of this metallic glass is 0.33 [9]. The elastic constant have been determined for a similar metallic glass, $Mg_{58}Cu_{30}Y_{12}$, from which we have $K/G$=2.4 and $\nu_{\text{Poisson}}$=0.32. By treating these two Mg-metallic glasses as the same one, the smaller values of $K/G$ and $\nu_{\text{Poisson}}$ correlate with the small value of $n$. The closely related metallic glass, $Mg_{65}Cu_{25}Gd_{10}$ has $K/G$=2.34 and $\nu_{\text{Poisson}}$=0.313, and $Mg_{65}Cu_{25}Tb_{10}$ has $K/G$=2.28 and $\nu_{\text{Poisson}}$ =0.309 [9]. The values of the elastic constants are about the same for all three Mg-based metallic glasses.

Finally, by considering the data of the Ce- and Mg- BMG in conjunction with the ductile $Zr_{46.75}Ti_{8.25}Cu_{7.5}Ni_{10}Be_{27.5}$ (VIT4), $Pd_{40}Ni_{40}P_{20}$, and $La_{70}Ni_{15}Al_{15}$, the correlation of $K/G$ (or $\nu_{\text{Poisson}}$) with $n$ or $1/\tau_\beta(T_g)$ is strengthened.

## VI Correlation of ductility with relaxation strength $\Delta_\beta$ of the $\beta$-relaxation on annealing

Mention has been made in the Introduction that the BMG becomes less ductile or more brittle by annealing, as demonstrated by the correlation of fracture energy $G_c$ with ratio $K/G$ as $Zr_{41}Ti_{14}Cu_{12.5}Ni_{10}Be_{22.5}$ (Vitreloy 1) is annealed up to 24 hours at 623 K [2]. Another study is the change in elastic and vibrational properties of $Zr_{46.75}Ti_{8.25}Cu_{7.5}Ni_{10}Be_{27.5}$ (Vit4) with annealing time at 523 ± 0.1 K which is 127 K below its $T_g$ [48]. The Poisson's ratio $\nu_{\text{Poisson}}$ decreased with annealing time, although the change is only 1% probably due to low annealing temperature.

Annealing study by isochronal dynamic mechanical spectroscopy at 1 Hz at 423 K of $La_{60}Ni_{15}Al_{25}$ showed, on increasing annealing time up to 24 hours, not only the α-relaxation shifts to higher temperatures as expected but also the $\beta$-relaxation, and the relaxation strength of the latter, $\Delta_\beta$, decreases monotonically [49]. The behavior of the $\beta$-relaxation on annealing was found in other metallic glasses [9], as well as in polymeric [50-53] and small molecular glass-formers [54-56]. Combining this change of the $\beta$-relaxation property on annealing with that of ductility (from fracture energy $G_c$), $K/G$, and Poisson's ratio $\nu_{Poisson}$, the existence of the correlation $\Delta_\beta$ and $\tau_\beta$ of the $\beta$-relaxation with ductility, $K/G$, and $\nu_{Poisson}$ on annealing is clear.

**VII. Derivation of the correlation between ductility and $K/G$ (or $\nu_{Poisson}$)**

The main thrust and goal of this paper is to understand from the more basic and microscopic dynamic properties of metallic glasses the reason why there is a correlation between the macroscopic properties of ductility/plasticity and the ratio $K/G$ or the Poisson's ratio $\nu_{Poisson}$. The observation of the ductile-brittle transition follows similar time-temperature scaling relationship of the activation of the $\beta$-relaxation has been elaborated in Section V. This remarkable correlation obviously indicates that the activation of the $\beta$-relaxation is responsible for the ductility of BMG. Moreover, by comparing the dynamic properties of the ductile La-, Zr- and Pd- BMG with the brittle Ce-BMG and Mg-BMG, established is the correlation of $K/G$, $\nu_{Poisson}$, and ductility with the non-exponentiality parameter of the structural α-relaxation, $n$, and the ratio, $\tau_\alpha/\tau_\beta$, for fixed value of $\tau_\alpha$. The reduction of ductility on annealing is accompanied by shift of $\tau_\beta$ to longer times and decrease of the relaxation strength, $\Delta_\beta$, discussed in Section VI. Thus these correlations between ductility, a macroscopic nonlinear mechanical property, and the microscopic α- and β-relaxation properties, constitute the first step towards the goal of relating ductility to $K/G$ and the Poisson's ratio $\nu_{Poisson}$ via microscopic considerations.

Besides Sections V and VI, other correlations have been established in Sections II-IV between the parameters selected from the inter-particle potential (anharmonicity), the reciprocal or the complement of the effective Debye-Waller factor, $1/f_0(T)$ and $(1-f_0)$ respectively, the non-exponentiality parameter of the α-relaxation ($n$), the ratio ($\tau_\alpha/\tau_\beta$) for fixed value of $\tau_\alpha$, the reciprocal of the strength of the β-relaxation $1/\Delta_\beta$, and $K/G$ or $\nu_{Poisson}$. We now utilize the totality of the correlations in Sections II-VI to link ductility together with $K/G$ or $\nu_{Poisson}$ via the microscopic dynamic properties, and hence derive the correlation between the elastic constants and ductility. One way to demonstrate this link is by the construction of Table 1, where the various correlations are shown. The link can be gleaned from the connectivity of the entries that enables ductility to reach $K/G$ or $\nu_{Poisson}$ by steps like descending on a staircase. All the steps traversed are microscopic quantities or properties, and hence the explanation of the correlation between ductility and $K/G$ or $\nu_{Poisson}$ given is based on microscopic dynamics of metallic glasses.

**VIII Discussions and Conclusions**

The empirical correlation between ductility and $K/G$ or $\nu_{Poisson}$ involving two macroscopic mechanical properties of bulk metallic glasses (BMG) seems to be universal. If the origin of the correlation is understood based on microscopic dynamic properties, the link of Poisson's ratio to basic physics governing the properties of BMG can be gained. The correlation also can be used to guide the fabrication of new BMG with improve plasticity for applications. Effort in this direction undertaken in the past is limited to exploring another correlation of $K/G$ or $\nu_{Poisson}$ [28,57] with $m$, as well as the fracture energy $G_c$ with $m$ [58], the steepness index or fragility [59,60]. The latter is derived from the temperature dependence of viscosity $\eta$ data in the equilibrium liquid state by

$$m = d\log\eta/d(T_{ref}/T)|_{T=T_{ref}} \qquad (6)$$

where $T_{ref}$ is usually defined as the temperature at which $\eta=10^{11}$ or $10^{12}$ Pa.s. Although the correlation between $K/G$ or $\nu_{Poisson}$ and $m$ is interesting and theoretical rationalization has been given [61], it comes short of explaining the correlation between ductility and *K/G* or $\nu_{Poisson}$. This shortcoming seems natural because there is no reason to expect *m*, an equilibrium liquid state property, to have direct and straightforward connection with either ductility or $\nu_{Poisson}$, which are properties of the glassy state.

Also the correlation runs into problem when considering the change of both quantities in the *same* glass-former on either isothermal annealing or on slow heating as done by Johari [62]. As a glass structurally relaxes on either isothermal annealing or on slow heating and its state corresponds to a higher value of the ratio of the activation energy to temperature, but both *K/G* and $\nu_{Poisson}$ decrease. Johari pointed out that this is the opposite of a correlation between *K/G* and $E_a/T_g$ or *m* that has been obtained by combining the data of different accuracy for chemically different liquids. We have a simpler argument to show the limitation in understanding the correlation between *K/G* or $\nu_{Poisson}$ and ductility. Already shown by Lewandowski et al. [2] annealing induces embattlement, as shown by the monotonic decrease of $G_c$ (or ductility) which is accompanied by a corresponding decrease of *K/G* or $\nu_{Poisson}$. That is $G_c$ and *K/G* or $\nu_{Poisson}$ remain correlated as the annealing progresses. Moreover, the correlation is the same as that found when comparing different BMG. However, *m* is a characteristic of the equilibrium liquid state and it is the same independent of the duration of annealing, and naturally *m* stops correlating with the decrease of $G_c$ and *K/G* on annealing.

Changes of *m* and *K/G* or $\nu_{Poisson}$ on progressively varying slightly the composition from $Cu_{50}Zr_{50}$ to $(Cu_{50}Zr_{50})_{96}Al_4$, and to $(Cu_{50}Zr_{50})_{90}Al_7Gd_3$ are interesting [9,63]. In the order of introducing the three BMG here, it was found that *m* decreases from 62 to 40, and to 30, while $\nu_{Poisson}$ increases from 0.395 to 0.369, and to 0.373. The trends of the changes of *m* and $\nu_{Poisson}$ are opposite to the correlation between *m* and $\nu_{Poisson}$ obtained by combining the data of different accuracy for chemically different BMG. There are other examples of violations of the correlation between $m$ and $\nu_{Poisson}$ worthwhile pointing out. For example, the brittle $Mg_{65}Cu_{25}Y_{10}$ has $m=41$ from [64], but $K/G=2.3$, and $\nu_{Poisson}=0.305$ given in Table 13 of [9]. The similar $Mg_{65}Cu_{25}Tb_{10}$ has $K/G=2.3$, $\nu_{Poisson}=0.309$, and $G_c=0.07$ kJ/m$^2$ [9]. On the other hand, the more ductile $Zr_{46.75}Ti_{8.25}Cu_{7.5}Ni_{10}Be_{27.5}$ (Vit4) has larger $K/G=3$, and $\nu_{Poisson}=0.35$, and yet its $m=38$ from [9,65] and $m=44$ from [9,66] are comparable to $m=0.41$ of $Mg_{65}Cu_{25}Y_{10}$. The fragility index $m$ of $Mg_{65}Cu_{25}Y_{10}$ is nearly the same or even larger than Vit4. This can be objectively deduced from the plot of log$\eta$ vs. $T_g/T$ of the two BMG in Fig.10 of [67].

Our explanation of the correlation between ductility and $K/G$ or $\nu_{Poisson}$ takes a different route. Ductility is linked to the secondary β-relaxation in the glassy state, and its relaxation time $\tau_\beta$ and strength is governed by the non-exponentiality parameter $n$ of the structural α-relaxation and its relaxation time $\tau_\alpha$. On the other end, $K/G$ or $\nu_{Poisson}$ is linked to the effective Debye-Waller factor (or non-ergodicity parameter) $f_0$, which in turn is related to $n$ since both quantities are determined by the anharmonicity of the inter-particle potential. From the links that ductility and $\nu_{Poisson}$ separately have with the parameters of the dynamics, the correlation found empirically between them is given an explanation based on microscopic considerations.

**Table 1.** The various correlations from Sections II-VI are entered in rows. They are arranged in such a way to connect ductility (at the top left corner) to the elastic constant, *K/G* or ν, (at the bottom right side) through their individual correlation with microscopic dynamic properties (lightly shaded). Like walking down a staircase, one can start from ductility to reach *K/G* or ν by steps all linked to microscopic dynamic properties. In this way, the correlation between ductility and the elastic constants is given an explanation based on microscopic dynamics of the BMG. The word "anharmonicity" is shortened to "AH" in the table.

| Ductility | | | | | | | Property | Source |
|---|---|---|---|---|---|---|---|---|
| Ductility | $\tau_\beta$ $E_\beta$ | | | | | | From the ductile-brittle transition vs. β-relaxation | [15] |
| | $\Delta_\beta$ | | | | | | From Annealing | [9,49] |
| Ductility | | $\tau_\alpha/\tau_\beta$ | $n$ | | | *K/G* ν | Ductile La-, Zr- and Pd- BMG vs. brittle Ce- and Mg- BMG | [9,12] |
| | | $\tau_\alpha/\tau_\beta$ | $n$ | AH | | | From the Coupling Model | [12,22-25] |
| | | | $n$ $n$ | AH | $1/f_0$ or $(1-f_0)$ NCL | | From simulations From experiments | [12,13] [30,37,38] |
| | | | | | $1/f_0$ or $(1-f_0)$ | *K/G* ν | From Eq.(3) | [26] |

**Acknowledgment**

The work performed at the Institute of Physics, Chinese Academy of Sciences, Beijing was supported by the NSF of China (51271195 and 11274353). We thank Prof. V. N. Novikov for stimulating discussion on the relation between the effective Debye-Waller factor and elastic constants.

**References**

[1] H. S. Chen, J. T. Krause, and E. Colemen. J Non-Cryst Solids **18,** 157 (1975).

[2] J. J. Lewandowski, W. H. Wang, and A. L. Greer, Philos Mag Lett **85**, 77 (2005).

[3] W. H. Wang, J. Appl. Phys. **99**, 093506 (2006).

[4] J. Schroers, and W. L. Johnson, Phys. Rev. Lett. **93**, 255506 (2004).

[5] W. H. Wang, R. J. Wang, W. T. Yang, B. C. Wei, P. Wen, D. Q. Zhao, M.X. Pan, J. Mater. Res. **17**, 1385 (2002).

[6] X. J. Gu, A. G. McDermott, S. J. Poon, and G. J. Shiflet, Appl. Phys. Lett. **88**, 211905 (2006). X. J. Gu, S. J. Poon, G. J. Shiflet, and M. Widom, Acta. Mater. **56**, 88 (2008).

[7] P. Yu, and H. Y. Bai, Mater.Sci.Eng.A **485**, 1 (2008).

[8] A. Castellero, D. I. Uhlenhaut, B. Moser, and J. Loffler, Philos. Mag. Lett. **87**, 383 (2007).

[9] W.H. Wang, Prog. Mater. Sci. **57**, 487 (2012).

[10] T. Ichitsubo, E. Matsubara, T. Yamamoto, H. S. Chen, N. Nishiyama, J. Saida, and K. Anazawa, Phys. Rev. Lett. **95**, 245501 (2005).

[11] H.B. Yu, Z. Wang, W.H. Wang, H.Y. Bai, J. Non-Cryst. Solids, **358,** 869 (2012).

[12] K. L. Ngai, Z. Wang, X. Q. Gao, H. B. Yu, and W. H. Wang, J. Chem. Phys. **139**, 014502 (2013).

[13] H. B. Yu, K. Samwer, Y. Wu, and W. H. Wang, Phys. Rev. Lett. **109**, 095508 (2012).

[14] H. B. Yu, W.H. Wang, H.Y. Bai, Y. Wu, and M.W. Chen, Phys. Rev. B **81**, 220201 (2010).

[15] H. B. Yu, X. Shen, Z. Wang, L. Gu, W. H. Wang, and H.Y. Bai, Phys. Rev. Lett. **108**, 015504 (2012).

[16] A. S. Argon, Acta. Mater. **27,** 47 (1979).

[17] B. Zhang, D. Q. Zhao, M. X. Pan, W. H. Wang, and A. L. Greer, Phys. Rev. Lett. **94**, 205502 (2005).

[18] A. F. Yee, and S. A. Smith, Macromolecules **14**, 54 (1981).

[19] E. W. Fischer, G. P. Hellmann, H. W. Spiess, F. J. Horth, U. Ecarius, and M. Wehrle, Makromol. Chem., Suppl. **12**, 189 (1985).

[20] C. Xiao, J. Wu, L. Yang, A. F. Yee, L. Xie, D. Gidley, K. L. Ngai, and A. K. Rizos, Macromolecules **32***,* 7921 (1999).

[21] D. Pan, A. Inoue, T. Sakurai, and M. W. Chen, PNAS **105**, 14769 (2008).

[22] K. L. Ngai, Comment Solid State Phys. **9**, 127 (1979).

[23] K. Y. Tsang, and K. L. Ngai, Phys. Rev. E **54**, R3067 (1996), *ibid*. **56**, R17 (1997).

[24] K. L. Ngai, and K. Y. Tsang, Phys. Rev. E **60**, 4511 (1999).

[25] K. L. Ngai, *Relaxation and Diffusion in Complex Systems*, Springer, New York (2011).

[26] V. N. Novikov, and A. P. Sokolov, Nature **431,** 961 (2004).

[27] W. Götze and L. Sjögren, Rep. Prog. Phys. **55**, 241 (1992).

[28] K. L. Ngai, J. Chem. Phys. **109**, 6982 (1998).

[29] K. L. Ngai, Z. Wang, X. Q. Gao, H. B. Yu, and W. H. Wang, J. Chem. Phys. **139**, 014502 (2013).

[30] K. L. Ngai, Philos. Mag. **84**, 1341 (2004).

[31] T. A. Litovitz and C. M. Davis in *Physical Accoustics* Vol. **11 A,** ed. W. P. Mason (New York: Academic) pp 282-349 (1965).

[32] Patrice Bordat, Frederic Affouard, Marc Descamps, and K. L. Ngai, Phys. Rev. Lett. **93**, 105502 (2004).

[33] Patrice Bordat, Frederic Affouard, and Marc Descamps, J. Non-Cryst. Solids **353**, 3924 (2007).

[34] R. Böhmer, K.L. Ngai, C.A. Angell, D.J. Plazek, J. Chem. Phys. **99**, 4201 (1993).

[35] T. Scopigno, G. Ruocco, F. Sette, and G. Monaco, Science **302**, 849 (2003).

[36] S. Capaccioli, M. Shahin Thayyil, and K. L. Ngai, J. Phys. Chem. B **112,** 16035 (2008).

[37] K. L. Ngai , J. Habasaki, C. Léon, and A. Rivera, Z. Phys. Chem. **219,** 47 ((2005).

[38] K. L. Ngai, J.Non-Cryst.Solids **275,** 7 (2000).

[39] J. M. Pelletier, B. Van de Mortéle, and I. R. Lu, Mater. Sci. Eng., A **336**, 190 (2002).

[40] P. Rösner, K. Samwer, and P. Lunkenheimer, Eur. Phys. Lett. **68**, 226 (2004).

[41] J. Hachenberg, and K. Samwer, J. Non-Cryst. Solids, **352**, 5110 (2006).

[42] K. L. Ngai, J. Non-Cryst. Solids, **352**, 404 (2006).

[43] K. L. Ngai and M. Paluch, J. Chem. Phys. **120**, 857 (2004).

[44] S. Capaccioli, M. Paluch, D. Prevosto, L.-M. Wang, and K. L. Ngai, J. Phys. Chem. Lett. **3**, 735 (2012).

[45] B. Zhang, D. Q. Zhao, M. X. Pan, W. H. Wang, and A. L. Greer, Phys. Rev. Lett. **94**, 205502 (2005). Wei Hua Wang, J. Appl. Phys. **110**, 053521 (2011).

[46] X.F. Liu, B. Zhang, P. Wen, and W.H. Wang, J. Non-Cryst.Solids, **352**, 4013 (2006).

[47] B. Ruta, Y. Chushkin, G. Monaco, L. Cipelletti, E. Pineda, P. Bruna, V. M. Giordano, and M. Gonzalez-Silveira, Phys. Rev. Lett. **109**, 165701 (2012).

[48] P. Wen, G. P. Johari, R. J. Wang, and W. H. Wang, Phys Rev B **73**, 224203 (2006).

[49] Z. Wang, Ph.D. thesis, Institute of Physics, Chinese Academy of Science (2012), and to be published.

[50] G. P. Johari, J. Chem. Phys. **77**, 4619 (1982).

[51] M. Beiner, F. Garwe, K. Schröter, and E. Donth, Polymer **35**, 4127 (1994).

[52] A. Alegria, L. Goitiandia, I. Telleria, and J. Colmenero, Macromolecules, **30**, 3881 (1997).

[53] R. Casalini, and C. M. Roland, Phys. Rev. Lett. **102**, 035701 (2009).

[54] G. P. Johari, G. Power, and J. K. Vij, J. Chem. Phys. **117**, 1714 (2002).

[55] D. Prevosto, S. Capaccioli, M. Lucchesi, P. A. Rolla, and K. L. Ngai, J. Chem. Phys. **120**, 4808 (2004).

[56] P. Lunkenheimer, R. Wehn, U. Schneider, and A. Loidl, Phys. Rev. Lett. **95,** 055702 (2005).

[57] See also Spyros N. Yannopoulos, and G. P. Johari, Nature **442,** E7 (2006).

[58] G. N. Greaves, A. L. Greer, R. S. Lakes, and T. Rouxel, Nature Mater. **10**, 823 (2011).

[59] R. Böhmer, K.L. Ngai, C.A. Angell, and D.J. Plazek, J. Chem. Phys. **99**, 4201 (1993).

[60] In considering correlation of a property or quantity with $m$ of different glass-formers, it is important to strictly fix the value of $\eta(T_{ref})$. It is also important to have real data of $\eta$ over a range of temperature near $T_{ref}$ to obtain reliable value of $m$. Since metallic alloys tend to crystallize, viscosity data near $T_{ref}$ of many BMG are not measured, and values of $m$ are determined by the Vogel-Fulcher-Tammann fits to data obtained at higher temperatures.

should not be taken seriously. Such values of *m* are unreliable and have appeared in papers including tht by T. Komatsu in J. Non-Cryst.Solids, **185**, 199 (1995), and by J.H. Na, et al. in J. Mater. Res. **23**, 523 (2008). In contrast, the values of *m* collected in the paper, O.N. Senkov in Phys.Rev. B **76**, 104202 (2007), are reliable because they are obtained from viscosity data taken over a range of temperature near $T_{ref}$.

[61] T. Egami, Intermetallics **14**, 882 (2006).

[62] G.P. Johari, Philos. Mag. **86**, 1567 (2006).

[63] Y. Li, H. Y. Bai, W. H. Wang, and K. Samwer, Phys Rev B **74**, 052201 (2006).

[64] Eloi Pineda, Pere Bruna, Beatrice Ruta, Marta Gonzalez-Silveira, and Daniel Crespo, Acta Materialia **61**, 3002 (2013).

[65] P. Wen, D. Q. Zhao, M. X. Pan, W. H. Wang, Y. P. Huang, and M. L. Guo, Appl. Phys. Lett. **84,** 2790 (2004).

[66] R. Bruning, and K. Samwer, Phys. Rev. B **46**, 11318 (1992).

[67] R. Busch, W. Liu, and W. L. Johnson, J. Appl. Phys. **83**, 4134 (1998).